\documentclass[twocolumn]{jpsj2}

\def\runauthor{Akira {\sc Oguri}}

\title{ 
Fermi liquid theory for the nonequilibrium 
Kondo effect at low bias voltages 
}

\author{Akira {\sc Oguri}}

\inst{
          Department of Material Science,
          Osaka City University, 
          Sumiyoshi-ku, Osaka 558-8585,
          Japan 
}

\recdate{\today}

\abst{
In this report, 
we describe a recent development 
in a Fermi liquid theory for the Kondo effect in quantum dots 
under a finite bias voltage $V$. 
Applying the microscopic theory of Yamada and Yosida  
to a nonequilibrium steady state,
we derive the Ward identities for the Keldysh Green's function,
and determine the low-energy behavior of the differential 
conductance  $dI/dV$ exactly up to terms of order  $(eV)^2$ 
for the symmetric Anderson model.
These results are deduced from the fact 
that the Green's function at the impurity site 
is a functional of a nonequilibrium distribution $f_{\text{eff}}(\omega)$,  
which at $eV=0$ coincides with the Fermi function.
Furthermore, we provide an alternative description of
the low-energy properties using  
a renormalized perturbation theory (RPT).
In the nonequilibrium  state the unperturbed part of 
the RPT is determined by the renormalized free quasiparticles, 
the distribution function of which is given by  $f_{\text{eff}}(\omega)$.
The residual interaction between the quasiparticles $\widetilde{U}$,
which is defined by the full vertex part at zero frequencies, 
is taken into account by an expansion in the power series of $\widetilde{U}$.
We also discuss the application of the RPT 
to a high-bias region beyond the Fermi-liquid regime.
}

\kword{Kondo effect, Fermi liquid, 
 Nonequilibrium, Keldysh formalism, Anderson model, Quantum dot 
 }

\begin{document}

\sloppy

\maketitle

\section{Introduction}
\label{sec:Introduction}

The Kondo effect\cite{Kondo} in quantum dots has been 
an active research field over a decade.
The early prediction about a characteristic 
gate-voltage dependence of 
the linear-response conductance\cite{NgLee,GlatzmanRaikh,Kawabata} 
has been confirmed experimentally in semiconductor devices,
\cite{Goldharber,Cronenwett,Simmel,VanDerWiel}
and other new features of the Kondo physics 
are also being studied extensively in various situations,  
such as an AB ring, Josephson junction, ferromagnetic leads, etc.

The equilibrium and linear-response properties of a single quantum dot 
connected to normal leads can be explained basically 
based on the knowledge of the Kondo problem 
in dilute magnetic alloys,\cite{Hewson} 
although there exists some differences 
in experimental geometry (configuration)
between the impurity in magnetic alloys and 
quantum dots in semiconductor devices.
Therefore, the low-energy properties can be described 
by the local Fermi liquid theory,
\cite{Nozieres,YamadaYosida,Shiba}  
and the nonperturbative approaches developed 
for the alloys, such as the quantum Monte Carlo\cite{ao6,Sakai}
 and numerical renormalization group (NRG) methods,\cite{Izumida}
can be applicable to the quantum dots.
Particularly, the NRG has been used successfully to calculate 
the linear-response conductance of the quantum dots.\cite{Izumida}

The nonlinear transport under a finite bias voltage $V$, however, 
is still not fully understood, despite of a number of theoretical efforts.
\cite{WM,HDW2,TheoriesII,Schiller,KNG,KSL,ao10,ao13,FujiiUeda} 
Among a variety of aspects of the nonequilibrium properties,
in this report we focus our attention mainly on the low-energy properties. 
Specifically, we describe the Fermi-liquid behavior 
of the first nonlinear term of the differential conductance $dI/dV$ 
using the Ward identities, which is derived by applying  
the perturbation theory in the Coulomb interaction $U$ 
of Yamada and Yosida to the Keldysh Green's function. 
We show that the low-energy asymptotic form of 
the order $U^2$ self-energy\cite{HDW2} is 
essentially retained in all orders in $U$, 
and the contributions of the higher-order terms 
are absorbed into the coefficients which  can be written 
in terms of the {\em local-Fermi-liquid\/} parameters such as 
the width of the Kondo resonance $\widetilde{\Delta}$ 
and Wilson ratio $R$.\cite{ao10} 
The proof was provided previously in ref.\ \citen{ao10}.
In the present report, however, we give another derivation, 
using the property of the impurity 
Green's function $\mbox{\boldmath $G$}(\omega)$ 
as a functional of a nonequilibrium distribution 
function $f_{\text{eff}}(\omega)$, 
through which the dependence of $\mbox{\boldmath $G$}(\omega)$
on $eV$ and $T$ arises.
This property also allows us to deduce 
some exact results in the limit of large $eV$.\cite{ao13} 
In the present report, we re-examine the low-energy properties 
with an emphasis on this  aspect of the Green's function as a functional.

We also present an alternative description of   
the low-voltage Fermi-liquid behavior  
using the renormalized perturbation theory (RPT).\cite{HewsonRPT}
The unperturbed Green's function of the RPT 
in the Keldysh formalism consists of  
the propagators of the free quasiparticles, 
which are determined by  
the renormalized resonance of the width $\widetilde{\Delta}$ 
and the nonequilibrium distribution $f_{\text{eff}}(\omega)$. 
To second order in the residual interaction $\widetilde{U}$, 
which is  defined in eq.\ (\ref{eq:renorm_parm}), 
it gives the exact low-energy  $(eV)^2$ coefficient of $dI/dV$. 
The higher order terms in $\widetilde{U}$  
determine the high-energy properties. 
It has recently been confirmed that in equilibrium a combination 
of the RPT and NRG gives an efficient way of calculating the temperature 
dependence of the susceptibility,\cite{HewsonNRG} 
so that the RPT seems to be one possible approach    
to the nonequilibrium properties beyond the Fermi-liquid regime.

 In \S \ref{sec:model},
 we describe the Keldysh formalism for the Anderson impurity   
in order to describe clearly the properties of 
  $\mbox{\boldmath $G$}(\omega)$ as 
  a functional of $f_{\text{eff}}(\omega)$.
In \S \ref{sec:low-voltage}, 
we consider the low-energy behavior of 
the self-energy at small $eV$ using the Ward identities, 
and give an exact low-energy 
expression of $dI/dV$ in the electron-hole symmetric case.
In \S \ref{sec:RPT}, the RPT is applied to the 
low-voltage Fermi-liquid regime, 
and the procedure of the perturbation expansion 
in  $\widetilde{U}$ in the Keldysh formalism is provided.
In the appendix, details of the Ward identities 
and properties of $\mbox{\boldmath $G$}(\omega)$ as 
  a functional of $f_{\text{eff}}(\omega)$ are given.

\section{Keldysh Formalism for the Anderson Model}
\label{sec:model}

We start with the single Anderson impurity connected 
to two leads at the left ($L$) and right ($R$):
\begin{align}
H \  &= \  H_{c} \, + \, H_{d} \, + \,  H_{\rm mix} \, + \, H_{U} \;,
\label{eq:H}
\\
 H_{c}  &=  
\sum_{\lambda=L,R} 
 \sum_{k\sigma} 
  \epsilon_{k \lambda}^{\phantom{0}}\,
         c^{\dagger}_{k \lambda \sigma} 
         c^{\phantom{\dagger}}_{k \lambda \sigma}
\,,
\label{eq:Hc}
\\             
 H_{d}  & =  \sum_{\sigma}   E_d \,  n_{d\sigma} \;, \quad 
 H_{U}  =  \frac{U}{2} \biggl(\sum_{\sigma}n_{d\sigma} -1\biggr)^2 \;,
\label{eq:Hd}
\\
 H_{\rm mix} \,  &=
 \sum_{\lambda=L,R} 
 \sum_{\sigma} v_{\lambda}^{\phantom 0}  \left(\,  
             d^{\dagger}_{\sigma}  \psi^{\phantom{\dagger}}_{\lambda \sigma}
      +\,   \psi^{\dagger}_{\lambda \sigma}  d^{\phantom{\dagger}}_{\sigma}
             \,\right)   ,
\label{eq:Hmix}
\end{align}
where $d_{\sigma}$ annihilates an electron with spin 
$\sigma$ at the dot,
$n_{d\sigma} = d^{\dagger}_{\sigma} d^{\phantom{\dagger}}_{\sigma}$,
and $E_d = \epsilon_d + U/2$.
We assume that the onsite potential $\epsilon_d$ is a constant 
independent of the bias voltage, and take   
the Fermi level at equilibrium $\mu$ to be 
the origin of the energy, i.e.,  $\mu = 0$.
In the lead at $\lambda$ ($= L,\, R$),
the energy spectrum is given by 
$\epsilon_{k \lambda} = \epsilon_{k}  + eV_{\lambda}$. 
To specify how the bias voltage $V$ is applied to each of the leads, 
we introduce a parameter $\alpha_{\lambda}$ such that   
$V_L = \alpha_L V$ and $V_R = - \alpha_R V$ with $\alpha_L + \alpha_R =1$. 
In eq.\ (\ref{eq:Hmix}),
 $v_{\lambda}$ is the tunneling matrix element 
between the dot and lead at $\lambda$, and 
 $\psi_{\lambda \sigma}^{\phantom{\dagger}} 
= \sum_k c_{k \lambda \sigma}^{\phantom{\dagger}} 
/\sqrt{N}$.  
 We will use units $\hbar=1$. 

In the thermal equilibrium, we know that the density matrix is given by 
 $\rho_{\rm eq}^{\phantom{0}} \propto \text{e}^{-\beta H}$, 
 and thus the Hamiltonian determines both the time evolution 
 and statistical weight. 
 However, in  a nonequilibrium steady state 
 the density matrix cannot be determined simply by $H$,
 and it depends on how the system has been driven to the steady state.  
 The Keldysh formalism has been used widely for this purpose 
 to determine the density matrix $\widehat{\rho}(t)$ 
 for nonequilibrium states.\cite{Keldysh,Caroli,Landau}

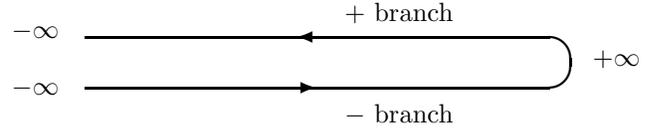
\begin{figure}[tb]
%% \begin{center}
\setlength{\unitlength}{0.75mm}
\hspace{-0.8cm}
\begin{picture}(120,20)(-5,0) 
\thicklines

\put(100,10){\oval(12,9)[r]} 

\put(60,5.5){\vector(1,0){1}}
\put(58.5,14.5){\vector(-1,0){1}}

\put(20,14.5){\line(1,0){80}}
\put(20,5.5){\line(1,0){80}}

\put(7,14){\makebox(0,0)[bl]{$-\infty$}}
\put(7,4){\makebox(0,0)[bl]{$-\infty$}}
\put(110,9){\makebox(0,0)[bl]{$+\infty$}}

\put(66,-1){\makebox(0,0)[bl]{$-$ branch}}
\put(66,17){\makebox(0,0)[bl]{$+$ branch}}

\end{picture}
 
\caption{The Keldysh contour for the time evolution.}
\label{fig:contour}
%% \end{center}
\end{figure}

The method uses the procedure of an adiabatic switching on,
which is described by the operator 
 ${\cal U}(t,t_0)  =  \text{T} \exp 
        [\,- \text{i} \int_{t_0}^t \text{d}t' \, 
        \widetilde{H}_2 (t') \,] $. 
Here, $\widetilde{O}(t)  \equiv  
      \text{e}^{\text{i} H_1 t}\, O\, \text{e}^{-\text{i} H_1 t}$ 
is an operator in the interaction representation with respect 
to $H_1$, which is a time-independent part of the total Hamiltonian 
$H(t)=H_1 + H_2\, \text{e}^{-\delta |t|}$.
In the interaction representation the density matrix  
defined by
$\widetilde{\rho}(t) \equiv  
 \text{e}^{\text{i} H_1 t} \, \widehat{\rho}(t)\, \text{e}^{-\text{i} H_1 t}$ 
can be rewritten in the form 
\begin{align} 
\widetilde{\rho}(t) \, = \, {\cal U}(t,-\infty) \, 
                         \widetilde{\rho}(-\infty) \, 
                         {\cal U}(-\infty,t) 
\label{eq:rho} \;,
\end{align}
where $\widetilde{\rho}(-\infty)$ represents 
the initial statistical weight.
The average value of 
a Heisenberg operator ${\cal O}_H(t) = {\cal U}(0,t) \, 
\widetilde{\cal O}(t) \, {\cal U}(t,0)$ is given by
\begin{align}
& \langle {\cal O}_H(t) \rangle \ \equiv \ 
            \mbox{Tr} \left[\,\widehat{\rho}(0) \,{\cal O}_H(t) \, \right]  
\nonumber
\\    
  &=  
  \text{Tr}
       \left[\, \widetilde{\rho}(-\infty) \,
      {\cal U}(-\infty,+\infty) \, 
      {\cal U}(+\infty,t) \, \widetilde{\cal O}(t) 
      \, {\cal U}(t,-\infty) \,\right].
\label{eq:average_H}
\end{align}
The stream of time seen in this expression 
is usually illustrated as the Keldysh contour shown  
in Fig.\ \ref{fig:contour}: 
the $+$ branch corresponds 
to the time evolution by the operator 
$ {\cal U}(-\infty,+\infty) =
 \widetilde{\text{T}} \exp 
[\, \text{i} \int_{-\infty}^{\infty} \text{d}t' 
\, \widetilde{H}_2 (t') \,] $, 
where $\widetilde{\text{T}}$ denotes 
the anti-time-ordering operator.
If one chooses $H_1$ to be bilinear, 
the Feynman-diagrammatic approach is applicable  
for the Green's functions defined by   
\begin{align} 
 G^{--}_{\sigma}(t) & \ = \  -\text{i} \,
\langle \text{T}\,
d^{\phantom{\dagger}}_{\sigma}(t)\, d^{\dagger}_{\sigma}(0)
\rangle  \;,
\label{eq:G^--}
\\
G^{-+}_{\sigma}(t) & \ = \     \text{i}\, 
\langle 
d^{\dagger}_{\sigma}(0) \, d^{\phantom{\dagger}}_{\sigma}(t) 
\rangle \;, 
\label{eq:G^-+}
\\
G^{+-}_{\sigma} (t) & \ = \  - \text{i} \,
\langle 
d^{\phantom{\dagger}}_{\sigma}(t) \, 
d^{\dagger}_{\sigma}(0)  
\rangle \;,
\label{eq:G^+-}
\\
G^{++}_{\sigma}(t) & \ = \ -\text{i} \,
\langle \widetilde{\text{T}}\,
d^{\phantom{\dagger}}_{\sigma}(t)\, d^{\dagger}_{\sigma}(0)
\rangle\;.
\label{eq:G^++}
\end{align} 
These functions are linearly dependent 
 $G^{-+}+G^{+-}=G^{--}+G^{++}$.
Furthermore, the retarded and advanced Green's functions 
can be written as $G^r = G^{--}-G^{-+}$ and $G^a = G^{--}-G^{+-}$, 
 respectively.

\subsection{Traditional Formulation}
\label{subsec:traditional}

\begin{figure}[tb]
%% \begin{center}
\setlength{\unitlength}{0.7mm}
%% \setlength{\unitlength}{0.6mm}
%% \begin{minipage}{1\linewidth}
\begin{picture}(105,49)(-56,-10)
\thicklines

\put(-45,0){\vector(1,0){110}}
\put(0,0){\vector(0,1){40}}

\put(-45,25){\line(1,0){30}}
\put(-15,15){\line(0,1){10}}
\put(-15,15){\line(1,0){40}}
\put(25,0){\line(0,1){15}}

\put(0,-5){\makebox(0,0){\large $0$}}
\put(-15,-5){\makebox(0,0){\large $\mu_R^{\phantom{0}}$}}
\put(27,-5){\makebox(0,0){\large $\mu_L^{\phantom{0}}$}}
\put(54,-7){\makebox(0,0){\large $\omega$ }}
\put(11,37){\makebox(0,0){\large $f_{\text{eff}}(\omega)$}}
\put(-4,26){\makebox(0,0){\large $1$}}
\put(47,9){\makebox(0,0)
   {\Large $\frac{\mathstrut \Gamma_L}{\mathstrut \Gamma_L+\Gamma_R}$}}

\thinlines

\put(32,10){\vector(0,-1){10}}
\put(32,10){\vector(0,1){5}}

\end{picture}
%% \end{minipage}
\caption{
The nonequilibrium distribution $f_{\text{eff}}(\omega)$
at $T=0$.}
\label{fig:distribution}
%% \end{center}
\end{figure}
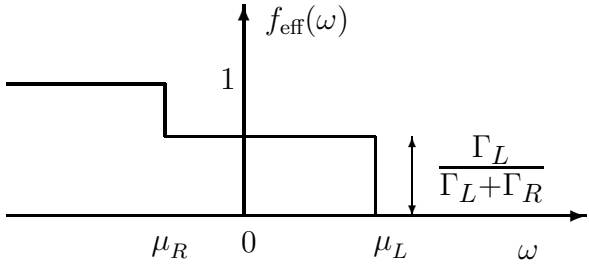

 To describe a nonequilibrium steady state 
 under a finite bias voltage, 
 Caroli {\em et al.\ }\cite{Caroli} has 
 introduced the initial statistical weight of the form 
\begin{equation}
\widetilde{\rho}(-\infty)
\, \propto  \, 
   \text{e}^{-\beta (\,H_d \,+\, H_c \,-\, \mu_L N_L \, -\, \mu_R N_R \,)} 
    \;, 
\label{eq:rho_ini}
\end{equation}
where 
$N_{\lambda} = \sum_{k\sigma} 
c^{\dagger}_{k \lambda \sigma} c^{\phantom{\dagger}}_{k \lambda \sigma}$.
The two chemical potentials, 
 $\mu_{L} \equiv eV_{L}$ and  $\mu_{R} \equiv eV_{R}$,
 are defined with respect to the isolated systems 
 described by $H_1 = H_{d} + H_{c}$,
  and  the remaining part $H_2 = H_{\rm mix} + H_U$ is  
  switched on adiabatically. 
Specifically, in the noninteracting case $U=0$,  the Green's functions 
can be written in the form
\begin{align}
 G_{0}^{--}(\omega) &=  
 \left[ 1-f_{\text{eff}}(\omega) \right]
     G_{0}^r(\omega)
     \, + \,  f_{\text{eff}}(\omega)\,G_{0}^a(\omega) 
\;,
\label{eq:G_0^--} 
\\
G_{0}^{-+}(\omega) & \,= \,  
-\,f_{\text{eff}}(\omega)
        \left[\,G_{0}^r(\omega)- G_{0}^a(\omega)\,\right] , 
\label{eq:G_0^-+}
\\
G_{0}^{+-}(\omega) & \,=  \,
\left[ 1-f_{\text{eff}}(\omega) \right]
        \left[\,G_{0}^r(\omega)- G_{0}^a(\omega)\,\right] , 
\label{eq:G_0^+-}
\\
 G_{0}^{++}(\omega) & \, = \, 
 -\left[ 1-f_{\text{eff}}(\omega) \right]
     G_{0}^a(\omega)
     \, - \,  f_{\text{eff}}(\omega)\,G_{0}^r(\omega) 
,
\label{eq:G_0^++}
\end{align}
where $G_{0}^{r}(\omega)  =  
  \bigl[\, \omega - E_d + \text{i} \Delta \,\bigr]^{-1}$, 
$G^a(\omega) = \{G^r(\omega)\}^*$,
and $\Delta = \Gamma_L + \Gamma_R$ with
 $\Gamma_{\lambda} = \pi \rho_{\lambda}^{\phantom{0}} v_{\lambda}^2$.
 We assume that the density of 
 states $\rho_{\lambda}^{\phantom{0}}(\omega) 
 = \sum_k 
\delta(\omega-\epsilon_{k \lambda}^{\phantom{0}})/N$ is a constant, 
and the band width is very large.
%% 
%%  $G^{++}(\omega)= - \{G^{--}(\omega)\}^*$.
%%
One important feature we see in eqs.\ 
(\ref{eq:G_0^--})--(\ref{eq:G_0^++}) is that
all the information about the nonequilibrium distribution
%% which must depend on $eV$ and $T$  
is contained in the distribution function,\cite{HDW2} 
\begin{equation}
f_{\text{eff}}(\omega) \, = \, 
{  f_L(\omega) \,\Gamma_L 
  + f_R(\omega) \,\Gamma_R
  \over 
 \Gamma_L +\Gamma_R } \;.
\label{eq:f_0}
\end{equation}
Here  $f_{\lambda}(\omega) 
= f(\omega - \mu_{\lambda}^{\phantom{0}})$,  
and $f(\omega)=[\,\text{e}^{\omega/T}+1\,]^{-1}$.
At $T=0$ the distribution function  $f_{\text{eff}}(\omega)$  has two steps, 
at $\omega=\mu_L$ and $\mu_R$,   
as shown in Fig.\ \ref{fig:distribution}. 
At $eV=0$, 
it coincides with the usual Fermi function $f(\omega)$.

The interacting Green's function $\mbox{\boldmath $G$}(\omega)$  
satisfies the Dyson equation, 
\begin{align}
& \{\mbox{\boldmath $G$}(\omega)\}^{-1} 
    =     
 \{\mbox{\boldmath $G$}_{0}(\omega)\}^{-1} 
-  \mbox{\boldmath $\Sigma$}(\omega) \;,
\label{eq:Dyson}
\\ 
& \mbox{\boldmath $G$}_{0} 
\, = \, 
\left[\, 
 \begin{matrix}
  G^{--}_{0} & G^{-+}_{0}   \cr
  G^{+-}_{0} & G^{++}_{0}  \cr  
 \end{matrix} 
 \, \right] , 
\quad
\mbox{\boldmath $\Sigma$} \, = \, 
 \left[ \,
  \begin{matrix}
   \Sigma^{--} & \Sigma^{-+}  \cr
   \Sigma^{+-} & \Sigma^{++}  \cr 
  \end{matrix}
 \,\right] .
\label{eq:Keldysh_Matrix}
\end{align}
Here $\mbox{\boldmath $\Sigma$}(\omega)$ is the self-energy due to $H_U$: 
the four elements are linearly 
dependent $\Sigma^{-+}+ \Sigma^{+-}= -\Sigma^{--}-\Sigma^{++}$,   
and we have also two extra relations in the $\omega$-space, 
$\Sigma^{a}(\omega)= \left\{\Sigma^{r}(\omega)\right\}^*$ 
and $\Sigma^{--}(\omega)= - \left\{\Sigma^{++}(\omega)\right\}^*$.
Using these relations, 
the retarded Green's function is written in the form
\begin{align}
G^{r}(\omega) \, = \, 
  {1 \over \omega -E_d + \text{i} \Delta - \Sigma^r(\omega) }
\end{align}
with $\Sigma^r = \Sigma^{--} + \Sigma^{-+}$.
The four elements of $\mbox{\boldmath $G$}(\omega)$ are also
written in the forms similar to eqs.\ (\ref{eq:G_0^--})--(\ref{eq:G_0^++}), 
for which $G^{r}_0$ and $G^{a}_0$ are replaced by the interacting ones 
and $f_{\text{eff}}(\omega)$ is replaced by a correlated distribution 
defined by 
\begin{align}
f_{\text{eff}}^U(\omega) & =
{ f_L(\omega) \,\Gamma_L 
  + f_R(\omega) \,\Gamma_R - 
  {\displaystyle \mathstrut 1 \over 
  \displaystyle \mathstrut 2\text{i}}\,\Sigma^{-+}(\omega)
  \over 
 \Gamma_L +\Gamma_R - \text{Im} \Sigma^r(\omega)} \;.
\label{eq:f_int}
\end{align}
This function was introduced 
by Hershfield {\em et al.}, and was studied  
using the order $U^2$ self-energy.\cite{HDW2}
Note that $\Sigma^{-+}(\omega)$ is pure imaginary,
and at $eV=0$ it takes the form 
 $\left.\Sigma^{-+}(\omega)\right|_{eV=0} = 
 2\text{i}f(\omega) 
 \text{Im}\left.\Sigma^{r}(\omega)\right|_{eV=0}$. 
Thus, in the nonequilibrium state, the distribution function   
$f_{\text{eff}}^U(\omega)$ generally depends on the interaction $U$,
while  in equilibrium  $eV=0$ it coincides with the Fermi function.

\subsection{Alternative Formulation}
\label{subsec:alternative}

As described in the above, 
the noninteracting Green's function for $H_0 \equiv H_c + H_d +H_{\rm mix}$ 
can be calculated analytically taking all contributions 
of the tunneling matrix element $H_{\rm mix}$ into account.
A question arises: 
do we always have to start with the isolated systems to 
obtain eqs.\ (\ref{eq:G_0^--})--(\ref{eq:G_0^++})?
The answer is no. 
An alternative description was given by Hershfield.\cite{Hershfield} 
The basic idea is to assign the two different chemical potentials 
directly to the left- and right-moving scattering states 
which are written formally, using the Lippmann-Schwinger equation, as 
\begin{align} 
\gamma_{k \lambda \sigma}^{\dagger}
\ = \ 
c_{k \lambda \sigma}^{\dagger}
\,+\,
\frac{1}{\varepsilon_k - H_0 +\text{i} \delta}
\,H_{\rm mix} \,
c_{k \lambda \sigma}^{\dagger} \;,
\label{eq:L-S}
\end{align} 
where $\lambda = L,\, R$. The incident wave comes in  
the left for $\gamma_{kL\sigma}^{\dagger}$, 
and in the right for $\gamma_{kR\sigma}^{\dagger}$.  
These scattering states are the eigenstates,
by which $H_0$ can be diagonalized as   
\begin{align}
H_0 \, = \, \sum_{\lambda=L,R}\sum_{k\sigma} 
 \varepsilon_{k}^{\phantom{0}} 
 \gamma_{k\lambda\sigma}^{\dagger} 
 \gamma_{k\lambda\sigma}^{\phantom{0}} \;.
\label{eq:H0_Jost} 
\end{align} 
Note that generally the bound states and continuum states 
without the degeneracy (for the left and right movers) are present. 
Such states are not distinguished from the degenerate scattering states 
in eq.\ (\ref{eq:H0_Jost}) for simplicity.
With these scattering states,
the density matrix for $U=0$ can be expressed explicitly as\cite{Hershfield} 
\begin{equation}
\widehat{\rho}_0^{\phantom{0}}(0)
\, \propto  \, 
   \text{e}^{ -\beta (\,H_d \,+\, H_c \,+\, H_{\rm mix} 
  \, -\, \mu_L \mathcal{N}_L
  \,-\, \mu_R  \mathcal{N}_R
\, )} 
    \;, 
\label{eq:rho_0}
\end{equation}
where
$  \mathcal{N}_{\lambda\sigma} = \sum_{k} 
\gamma_{k\lambda\sigma}^{\dagger} 
 \gamma_{k\lambda\sigma}^{\phantom{0}}  
$. 
One can confirm that the noninteracting Green's functions 
eqs.\ (\ref{eq:G_0^--})--(\ref{eq:G_0^++}) 
can be calculated directly from eq.\ (\ref{eq:rho_0}).
Therefore, the Coulomb interaction  $H_U$ can be switched on 
starting from the connected system   
taking $\widehat{\rho}_0(0)$, given in eq.\ (\ref{eq:rho_0}), 
to be the initial statistical weight. 
It is carried out by using eq.\ (\ref{eq:average_H}) 
and redefining the initial condition as 
 $H_1 \Rightarrow H_0$, $H_2 \Rightarrow H_U$, 
 and  $\widetilde{\rho}(-\infty) \Rightarrow 
  \text{e}^{ -\beta (\,H_0 
  \, -\, \mu_L \mathcal{N}_L
  \,-\, \mu_R  \mathcal{N}_R
\, ) } 
$.

The perturbation series in $U$ of 
the nonequilibrium Green's 
function $\mbox{\boldmath $G$}$ 
can be generated automatically using the path 
integral representation,\cite{PathIntegral} 
\begin{align}
\mathcal{Z}  &\, =    
\int \!
{\cal D}\eta^{\dagger} {\cal D}\eta \  \text{e}^{\text{i} S}
\;,
\label{eq:ZJ}
\\
G^{\nu\nu'}_{\sigma}(t,t') &\, = \, 
{ -\text{i}\over \mathcal{Z} } 
\int \!
{\cal D}\eta^{\dagger} {\cal D}\eta \  \text{e}^{\text{i} S}
\,\eta_{\sigma \nu}^{\phantom{\dagger}}(t)
\,
\eta_{\sigma \nu'}^{\dagger}(t')
\;,
\label{eq:Green_path}
\end{align}
where $\eta_{\sigma \nu}^{\phantom{\dagger}}(t)$ 
is a Grassmann number for the branch $\nu$ ($=-,\,+$) in 
the Keldysh contour. 
The action $S$ is defined by 
\begin{align}
   S & \ = \   
   S_0
      \, + \, 
   S_U \;, 
  \label{eq:action_tot}
   \\
   S_0 & \,=\,  
   \sum_{\sigma} \int_{-\infty}^{\infty} \! \text{d}t\,\text{d}t'\ 
 \mbox{\boldmath $\eta$}_{\sigma}^{\dagger}(t)\, 
       \mbox{\boldmath $K$}_0(t,t')\, 
 \mbox{\boldmath $\eta$}_{\sigma}(t') 
 \;, 
 \label{eq:action_0} 
   \\
   S_U & \, = \, 
   -\, \frac{U}{2}\! \int_{-\infty}^{\infty} \! \text{d}t \,
 \Biggl[\,  
 \biggl(\, \sum_{\sigma}
 \eta_{\sigma -}^{\dagger}(t)\,
 \eta_{\sigma -}^{\phantom{\dagger}}(t)
 \,-\,1  \,\biggr)^2
\nonumber \\
 & \qquad \qquad \qquad  -   \, 
\biggl(\,\sum_{\sigma}
 \eta_{\sigma +}^{\dagger}(t)\,
 \eta_{\sigma +}^{\phantom{\dagger}}(t)
 \,-\,1  \,\biggr)^2  
 \,\Biggr]  ,
  \label{eq:action_U}  
\end{align}
where
 \begin{align}
    \mbox{\boldmath $K$}_0(t,t') & \,  =  
    \int_{-\infty}^{\infty} \! {\text{d}\omega \over 2 \pi}\,
    \left\{\mbox{\boldmath $G$}_0(\omega)\right\}^{-1} 
     \text{e}^{-\text{i}\omega (t-t')} \;,
\label{eq:K_0}
 \end{align}
and 
$
 \mbox{\boldmath $\eta$}_{\sigma}^{\dagger}(t)
 = \bigl(\, 
 \eta_{\sigma -}^{\dagger}(t) \,, \, 
 \eta_{\sigma +}^{\dagger}(t) \,\bigr) 
 $. 
 In the Keldysh formalism 
the perturbation expansion works 
with the real frequencies (or real times). 
Therefore, eq.\ (\ref{eq:Green_path})  shows that 
the dependence of $\mbox{\boldmath $G$}(\omega)$
on the bias voltage and temperature 
arises through $f_{\text{eff}}(\omega)$  
in the noninteracting 
Green's function $\mbox{\boldmath $G$}_0(\omega)$ 
which determines $S_0$ via eqs.\ (\ref{eq:action_0}) and (\ref{eq:K_0}).
Thus, the full Green's function $\mbox{\boldmath $G$}(\omega)$, 
can be regarded as a functional of $f_{\text{eff}}(\omega)$.
The precise form of the functional 
is obtained by expanding $\text{e}^{\text{i} S}$ in eq.\ (\ref{eq:Green_path})
in the power series of $U$, 
and substituting eqs.\ (\ref{eq:G_0^--})--(\ref{eq:G_0^++}) 
into every single $\mbox{\boldmath $G$}_0$'s in the series.
Therefore, the change in the self-energy,  
$\delta \mbox{\boldmath $\Sigma$}$, 
caused by a small variation 
in the distribution function,  $\delta \! f_{\text{eff}}$, 
can be expressed in the form 
\begin{align}
 \delta \mbox{\boldmath $\Sigma$}_{\sigma}(\omega)
& \, = \, 
 \sum_{\nu\nu'\sigma'} 
\int_{-\infty}^{\infty} \! {\text{d}\omega'}   
\biggl.
{
\delta \mbox{\boldmath $\Sigma$}_{\sigma}(\omega)
\over
\delta G_{0\sigma'}^{\nu\nu'}(\omega')
}
\biggr|_{\delta \! f_{\text{eff}}=0}  
 \delta G_{0\sigma'}^{\nu\nu'}(\omega') 
\nonumber \\
& \quad  + \ 
\Bigl[\, \mbox{higer order terms in $\delta \! f_{\text{eff}}$} \, \Bigr]
\; ,
\label{eq:functional_f1}
%% \\
\end{align}
where
\begin{align}
 \delta G_{0\sigma'}^{\nu\nu'}(\omega')
&\,=\, 
- \left[\,G_{0\sigma'}^r(\omega')- G_{0\sigma'}^a(\omega')\,\right]  
\,\delta \! f_{\text{eff}}(\omega') \;.
\label{eq:delta_G0}
\end{align}
The functional derivative 
 $\delta \mbox{\boldmath $\Sigma$}
 _{\sigma}/
\delta G_{0\sigma'}^{\nu\nu'}$ can be related 
to the vertex corrections in the Keldysh formalism.
The functional aspect discussed here  
is analogous to the functional approach 
of Luttinger and Ward.\cite{LuttingerWard} 
However, in eq.\ (\ref{eq:functional_f1}), 
the functional derivative is taken with respect to 
noninteracting Green's function. 
 At finite temperatures $T \neq 0$  
the distribution function $f_{\text{eff}}(\omega)$ dose not have 
the discontinuities, 
and thus it can be treated as a regular function in general discussions.
Nevertheless, the singularities appearing in the limit of $T \to 0$ 
play an important role, for instance, 
as we see in eq.\ (\ref{eq:Ward_2_Im_diagram}).

\section{Fermi-Liquid Behavior at Small Voltages}
\label{sec:low-voltage}

In equilibrium and linear-response regime,
the low-energy properties at $\omega$, $T \ll T_K$
can be described by the local Fermi liquid theory,\cite{Nozieres}
where $T_K$ is the Kondo temperature. 
The Fermi liquid theory can also describe  
the nonlinear response at small bias-voltages $eV \ll T_K$.\cite{ao10} 
Our proof uses the Ward identities\cite{YamadaYosida,Shiba,Yoshimori}
in the Keldysh formalism.
In this section, we describe the outline of the derivation 
of the identities, and then determine the low-voltage behavior 
of the differential conductance $dI/dV$ up to terms of order $(eV)^2$   
in the electron-hole symmetric case.

\subsection{Ward identities}

We first of all consider the behavior of 
 $\mbox{\boldmath $G$}_0(\omega)$ at small $eV$.
 The first derivative at $eV=0$ is written in the form 
\begin{align}
  &
 \!\!\!\!
 \left.
  {\partial  
 G_0^{\nu\nu'}(\omega) 
 \over
 \partial ({\sl e}V)} 
 \right|_{eV=0}
 = \,
  -\, \alpha 
 \left(
 {\partial \over \partial \omega}
 + {\partial \over \partial E_d}
 \right)
 G_{0:\text{eq}}^{\nu\nu'}(\omega)\,,
  \label{eq:derivative_1b}
\end{align}
where 
$\alpha \equiv (\alpha_L \Gamma_L - \alpha_R \Gamma_R)/ 
   (\Gamma_L+ \Gamma_R)$, 
   and the label \lq\lq eq" in the subscript 
   stands for the \lq\lq equilibrium", so that
 $
 G_{0:\text{eq}}^{\nu\nu'}
 \equiv 
 \bigl.G_{0}^{\nu\nu'}\bigr|_{eV=0}$. 
Owing to the properties of $f_{\text{eff}}(\omega)$, 
the differential coefficient with respect to $eV$ 
can be related to the equilibrium quantities 
in the right-hand side eq.\  (\ref{eq:derivative_1b}).  
From the discussions in \S \ref{sec:model}, 
the self-energy can also be regarded a functional 
of $\mbox{\boldmath $G$}_{0}(\omega)$. 
Thus, the differential coefficients of  
$\mbox{\boldmath $\Sigma$}(\omega)$ with respect to $eV$ can be 
calculated taking the derivative 
of $\mbox{\boldmath $G$}_{0}$'s appearing 
in the perturbation series in $U$, as described in the appendix.
Then, using eq.\ (\ref{eq:derivative_1b}), we obtain  
\begin{align}
\left.
{\partial  
\mbox{\boldmath $\Sigma$}(\omega) 
\over
\partial ({\sl e}V)}\right|_{eV=0} 
 &= \ 
 -\,
 \alpha  
\left(
{\partial \over \partial \omega}
+ {\partial \over \partial E_d}
\right)
\mbox{\boldmath $\Sigma$}_{\text{eq}}(\omega) 
\;,
\label{eq:derivative_self_1}
\\
\left.
{\partial^2  
\mbox{\boldmath $\Sigma$}(\omega) 
\over
\partial ({\sl e}V)^2}\right|_{eV=0} 
& \, = 
\  \alpha^2
\left(
{\partial \over \partial \omega}
 + {\partial \over \partial E_d}
\right)^2 
\mbox{\boldmath $\Sigma$}_{\text{eq}}(\omega) 
 \nonumber \\
 & \quad \ \  + \,    
{ \Gamma_L\,\Gamma_R 
  \over \left( \Gamma_L+ \Gamma_R \right)^2}
   \  \widehat{D}^2 
\mbox{\boldmath $\Sigma$}_{\text{eq}}(\omega) 
\;,
\label{eq:derivative_self_2_with_CII}
\end{align}
where 
$\mbox{\boldmath $\Sigma$}_{\text{eq}}(\omega) 
\equiv \left. \mbox{\boldmath $\Sigma$}(\omega) \right|_{eV=0}$.   
The operator $\widehat{D}^2$ acts on 
the noninteracting Green's functions in the perturbation series 
for $\mbox{\boldmath $\Sigma$}_{\text{eq}}(\omega)$,  
and it takes the second derivative 
$(\partial /\partial \omega' + \partial/ \partial E_d )^2$, as  
\begin{align}
&   \widehat{D}^2 
\mbox{\boldmath $\Sigma$}_{\text{eq},\sigma}(\omega)
 \   =    
 \nonumber \\
& \quad  \sum_{\nu\nu'\sigma'} 
\int \! {\text{d}\omega'}   
{
\delta \mbox{\boldmath $\Sigma$}_{\text{eq},\sigma}(\omega)
\over
\delta G_{0:\text{eq},\sigma'}^{\nu\nu'}(\omega')
} 
\left(
{ \partial 
  \over
  \partial \omega' }
+{ \partial 
  \over
  \partial E_d }
  \right)^2 G_{0:\text{eq},\sigma'}^{\nu\nu'}(\omega') .
\end{align}
Using these relations, 
the low-bias behavior of the self-energy 
can be deduced from the equilibrium quantities.

\begin{figure}[t]
\leavevmode 
%% \begin{center}
\hspace{0.5cm}
\begin{minipage}{0.72\linewidth}
\includegraphics[width=1.2\linewidth,
, clip, trim = 0cm 0cm 0cm 0cm]{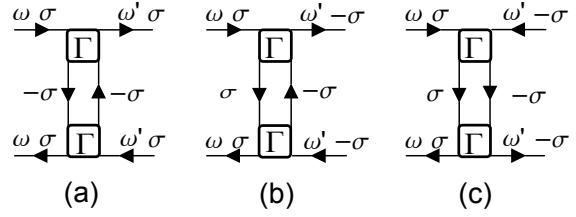}
\end{minipage}
\caption{The diagrams contribute to the singularities.} 
\label{fig:vertex}
%% \end{center}
\end{figure}

Specifically, at $T =0$ and $eV = 0$, 
the usual zero-temperature formalism  
is applicable for the causal Green's function  
defined with respect to the equilibrium ground state,
\begin{align}
G_{\rm eq}^{--}(\omega) \ =  \  
 G_{\rm eq}^{r}(\omega) \, \theta(\omega)
\,+\, G_{\rm eq}^{a}(\omega)\, \theta(-\omega)
\;,
\end{align} 
where $\theta (\omega)$ is the Heaviside step function.
Thus at $T=0$ the causal matrix-element 
in the right-hand side of eq.\ (\ref{eq:derivative_self_1})  
can be related to the vertex corrections,\cite{Yoshimori}
\begin{align}
& \left( {\partial \over \partial \omega} 
+ {\partial \over \partial E_d} \right)
 \Sigma_{{\rm eq},\sigma}^{--}(\omega) 
\ = \ 
\nonumber
\\
%% &= \sum_{\sigma'}\int {\text{d}\omega' \over 2\pi \text{i}}\, 
%% \Gamma_{\sigma\sigma';\sigma'\sigma}(\omega,\omega';\omega',\omega)
%% \,\delta G_{\sigma'}(\omega') 
%% \left\{-{\partial f(\omega') \over \partial \omega'} \right\}
%% \nonumber 
%% \\
& \qquad  \qquad  \quad -
\sum_{\sigma'} \,
\Gamma_{\sigma\sigma';\sigma'\sigma}(\omega,0;0,\omega) 
\,
A_{{\rm eq},\sigma'}(0)
\;, 
%% \\
\label{eq:Ward_1}
\end{align}
where $\Gamma_{\sigma\sigma';\sigma'\sigma}(\omega,\omega';\omega',\omega)$ 
is the vertex function for the causal Green's function
in the $T=0$ formalism, and 
 $A_{{\rm eq},\sigma}(\omega) 
= -\text{Im}\, G^{r}_{{\rm eq},\sigma}(\omega)/\pi$. 
Similarly, 
the causal element of $\widehat{D}^2 
\mbox{\boldmath $\Sigma$}_{\text{eq},\sigma}$ 
can be written as\cite{ao10}
\begin{align}
& \widehat{D}^2  \Sigma_{{\rm eq},\sigma}^{--}(\omega) 
\ = 
\nonumber
\\
%% & = 
%% \sum_{\sigma'}\int {\text{d}\omega' \over 2\pi \text{i}}\, 
%% \Gamma_{\sigma\sigma';\sigma'\sigma}(\omega,\omega';\omega',\omega)
%% \,\delta G_{\sigma'}(\omega') 
%% \left\{-{\partial^2 f(\omega') \over \partial {\omega'}^2} \right\}
%% \nonumber
%% \\
& \quad   
\sum_{\sigma'}  
{\partial \over \partial \omega'}\,  
\Bigl.
\Gamma_{\sigma\sigma';\sigma'\sigma}(\omega,\omega';\omega',\omega)
\,
A_{{\rm eq},\sigma'}(\omega')
 \,\Bigr|_{\omega'=0} \;.
\label{eq:D2_T0}
\end{align}

Eliashberg\cite{Eliashberg} has shown quite generally 
by using the Lehmann representation that 
the imaginary part of the vertex function has 
some singularities.\cite{Shiba,Yoshimori,ao11}
 For small frequencies, the singularities 
relevant to eqs.\ (\ref{eq:Ward_1}) and (\ref{eq:D2_T0}) 
 arise from the diagrams shown in Fig.\ \ref{fig:vertex}.  
The asymptotic form of the imaginary part of eq.\ (\ref{eq:D2_T0}) 
at small $\omega$ and $\omega'$ can be calculated as\cite{ao10}
\begin{align}
&   \ \  
\sum_{\sigma'}  
{\partial \over \partial \omega'}
\, \mbox{Im}\, 
\Gamma_{\sigma\sigma';\sigma'\sigma}(\omega,\omega';\omega',\omega)
\nonumber \\
& =    
-\,
\left|\Gamma_{\uparrow\downarrow;\downarrow\uparrow}(0,0;0,0)
\right|^2 
\nonumber \\
& \quad 
\times \, \mbox{Im} \Biggl[ \,
2 \int {d \omega'' \over 2\pi \text{i}}\,
G_{\rm eq}^{--}(\omega'') \,
{\partial  \over \partial \omega'} G_{\rm eq}^{--}(\omega -\omega'+ \omega'')
\nonumber \\
& \qquad \qquad 
+ \int {d \omega'' \over 2\pi \text{i}}\,
G_{\rm eq}^{--}(\omega'') \,
{\partial  \over \partial \omega'} G_{\rm eq}^{--}(\omega +\omega'-\omega'')
\,\Biggr]
\nonumber
\\
& = \,   
-\,
\pi \left\{A_{\rm eq}(0)\right\}^2\, \left| 
  \Gamma_{\uparrow\downarrow;\downarrow\uparrow}(0,0;0,0)\right|^2 \,
\nonumber\\
& \qquad \times 
\Bigl[\, 
- 2\,\mbox{sgn}(\omega'-\omega) \,
+ \,\mbox{sgn}(\omega' +\omega) \,
\Bigr] \;.
\label{eq:Ward_2_Im_diagram}
\end{align}
Here the first term in the last line corresponds to the 
contributions of the diagram (a) and (b), 
and the second term corresponds to that of the diagram (c).
Due to these singularities, 
the value of eq.\ (\ref{eq:Ward_2_Im_diagram}) 
in the limit of $\omega,\, \omega' \to 0$ depends 
on which frequency is taken first to be zero. 
Taking the limit for small frequencies carefully, 
the low-energy behavior of $\mbox{Im}\, \Sigma^r(\omega)$ 
are determined up to terms of order $\omega^2$, $(eV)^2$, and $T^2$,    
%%
%% \pagebreak
%%
\begin{align}
&\mbox{Im}\, \Sigma^r(\omega) 
  =    
   \, -\, 
  { \pi \over 2 } \left\{A_{\rm eq}(0)\right\}^3\, \left| 
  \Gamma_{\uparrow\downarrow;\downarrow\uparrow}(0,0;0,0)\right|^2
\,
\nonumber \\
%%
%%  { \pi A_{\rm eq}(0)\, \widetilde{\chi}_{\uparrow\downarrow}^2\over 2}\,
%%
 &  \times 
        \left[\,
            \left(\,\omega - 
              \alpha\, {\sl e}V\, 
              \right)^2 
              +  { 3\,\Gamma_L \Gamma_R 
                 \over \left( \Gamma_L + \Gamma_R \right)^2}
                \,(eV)^2 
              +(\pi T)^2  
           \,\right] . 
\label{eq:self_imaginary}
\end{align}
The result at equilibrium $eV=0$ has been 
provided by Yamada and Yosida,\cite{YamadaYosida} 
and it is extended to the nonequilibrium steady state here 
up to terms of order $(eV)^2$.
Note that we have not assumed the electron-hole symmetry so far.

%%
%%   \mathstrut
%%

\subsection{Results in the electron-hole symmetric case}

In this subsection we consider the low-energy behavior 
of $G^r(\omega)$  and $dI/dV$  using 
the result of $\mbox{Im}\, \Sigma^r(\omega)$ 
obtained in eq.\ (\ref{eq:self_imaginary}).
Specifically, we concentrate on  
the electron-hole symmetric case, where    
$\epsilon_d = -U/2$, $\Gamma_L = \Gamma_R$, and $\alpha_L=\alpha_R=1/2$. 
 In this case $A_{\rm eq}(0) = 1/(\pi\Delta)$, 
and the real part of the self-energy takes the form   
\begin{align}
\mbox{Re}\, \Sigma^r(\omega) &\,= \,\left( 1 - z^{-1} \right)
 \omega \, + \, O(\omega^3) \;,
\label{eq:re_S} 
\\
z &\,\equiv \,
\left(\,
1- \left.{\partial \Sigma_{\rm eq}^r(\omega) \over \partial \omega }
\right|_{\omega=0} \,\right)^{-1} \;.
%% \left [\,1- \left.\partial \Sigma_{\rm eq}^r(\omega)/\partial 
%% \omega \right|_{\omega=0} \,\right]^{-1} \;.
\label{eq:wave_remorm}
\end{align}
Thus, 
 $G^r(\omega)$ can be calculated  
exactly up to terms of order $\omega^2$, $T^2$ and $(eV)^2$ 
 using eq.\ (\ref{eq:self_imaginary}), 
\begin{align}
G^r(\omega) \simeq 
{z \over 
\omega + \text{i}\, \widetilde{\Delta} 
      +\,\text{i}\, 
 {\displaystyle \mathstrut \widetilde{U}^2 
 \over \displaystyle 
 2 \widetilde{\Delta} 
 \rule{0cm}{0.4cm} \mathstrut 
       (\pi \widetilde{\Delta})^2}
\left[ \, \omega^2 + 
{\displaystyle \mathstrut 3  \over \displaystyle \mathstrut 4}
\,({\sl e}V)^2 +(\pi T)^2
       \right]
       }         ,
 \label{eq:Gr_symmetric_case}
\end{align}
where the renormalized parameters are defined by 
\begin{align}
\widetilde{\Delta}\, &\equiv \,z \Delta \;, \qquad
\widetilde{U} \, \equiv\,  z^2\, 
\Gamma_{\uparrow\downarrow;\downarrow\uparrow}(0,0;0,0) \;.
\label{eq:renorm_parm}
\end{align}
The  order $U^2$ result of Hershfield {\em et al.}\cite{HDW2} 
can be reproduced from eq.\ (\ref{eq:Gr_symmetric_case})
replacing $\widetilde{U}$ by the bare Coulomb interaction $U$, 
and using the order $U^2$ result for 
the renormalization factor\cite{YamadaYosida,ZlaticHorvatic}
$z = 1 - (3-\pi^2/4) \,u^2 + \cdots$, where $u= U/(\pi \Delta)$.

\begin{figure}[t]
\leavevmode 
\begin{center}
\begin{minipage}{0.9\linewidth}
\includegraphics[width=1\linewidth]{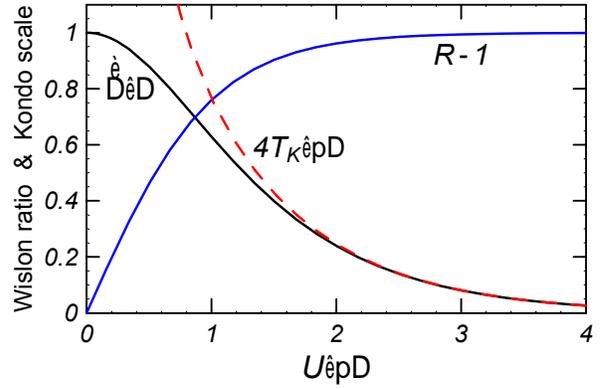}
\end{minipage}
\caption{The $U$ dependence of
$\widetilde{\Delta}/\Delta$ 
and 
$R-1$,  
which can also be interpreted as  
 $z$ and $\widetilde{U}/(\pi\widetilde{\Delta})$, respectively. 
These parameters were calculated using   
the Bethe ansatz solution summarized in ref.\ \citen{ZlaticHorvatic}. } 
\label{fig:Bethe}
\end{center}
\end{figure}

Thus, in the symmetric case  
the low-voltage behavior 
is characterized by the two parameters $\widetilde{\Delta}$  
and $\widetilde{U}$. 
These parameters  
are defined with respect to the equilibrium ground state,
for which the exact Bethe ansatz results exist  
\cite{KawakamiOkiji,WiegmanTsvelick,ZlaticHorvatic} 
as shown in Fig.\ \ref{fig:Bethe}. 
The width of the Kondo resonance $\widetilde{\Delta}$ decreases 
with increasing $U$, and 
for $u \gtrsim 2.0$ it is approximated well 
by the asymptotic form $\widetilde{\Delta} \simeq (4/\pi) T_K$,    
where the Kondo temperature is defined by 
\begin{align}
T_K \, = \, \pi \Delta \sqrt{u / (2 \pi)}\, \exp[-\pi^2 u/8 + 1/(2u)] \;.
\label{eq:T_K}
\end{align}
The Wilson ratio is usually defined by 
$R \equiv \widetilde{\chi}_s/\widetilde{\gamma}$,
where $\widetilde{\gamma}$ and $\widetilde{\chi}_s$ are  
the enhancement factors for the $T$-linear specific heat 
and spin susceptibility,  respectively.\cite{YamadaYosida}
Alternatively, it can be written in terms 
of $\widetilde{\Delta}$ and $\widetilde{U}$, as\cite{HewsonRPT} 
\begin{align}
R -1 \, = \,  \widetilde{U}/(\pi \widetilde{\Delta}) \;.
\label{eq:wilson_vs_renorm}
\end{align}
The Wilson ratio increases with $u$ from the noninteracting value  
$R=1$ to the strong-coupling limit value $R = 2$.
The charge excitations at the impurity site 
are still surviving for $u \lesssim 2.0$, 
and it makes the value of $R$ smaller than $2$.

The nonequilibrium current $I$ can be calculated from 
the retarded Green's function,\cite{MW} 
\begin{align}
%% \langle n_d  \rangle 
%% \, &= \, 
%%  2 \int_{-\infty}^{\infty} {\text{d}\omega \over 2\pi \text{i}}
%%  \,  G^{-+}(\omega) \;,  
%% \\
%%
 I\,  &  =\,  {2 {\sl e} \over h}  \int_{-\infty}^{\infty} 
 \! \text{d}\omega 
   \left[\, f_L - f_R \, \right] 
\frac{4\, \Gamma_L \Gamma_R}{\Gamma_R + \Gamma_L} 
 \left[\, - {\rm Im}\, G^r(\omega) \,\right] .
\label{eq:caroli}
\end{align}
Substituting     
eq.\ 
(\ref{eq:Gr_symmetric_case})  
into eq.\ (\ref{eq:caroli}),  
the differential conductance $dI/dV$ can be determined exactly  
up to terms of order $T^2$ and $(eV)^2$,  
%%
%% \begin{full}
\begin{align}
%% & A(\omega)
%% \ = \ {1\over \pi \Delta}  
%%   \Biggl [\, 1 \,
%%   - \,\left( 1 + {(R-1)^2 \over 2} \right) 
%%   \left({\omega \over \widetilde{\Delta}}\right)^2
%% \nonumber \\
%%    & \qquad 
%%  \,
%%    -\, {(R-1)^2 \over 2}  
%%                  \left( {\pi T \over \widetilde{\Delta}}\right)^2 
%%      -\, {3\, (R-1)^2 \over 8} 
%%                  \left( {eV \over \widetilde{\Delta}}\right)^2 
%%  + \cdots  \,\Biggr] ,             
%% \label{eq:Spectral}
%% \\
& {dI \over dV}  
\ = \ {2 e^2 \over h}  
  \Biggl[\, 1 \,
  - 
                \, { 1
                  + 2\, (R-1)^2 
                  \over 3} 
                 \left( {\pi T \over \widetilde{\Delta}}\right)^2
   \nonumber \\
 &  \qquad \qquad \qquad \;
 \,
      - 
                \, { 1
                  + 5\, (R-1)^2 
                  \over 4} 
                 \left( {eV \over \widetilde{\Delta} } \right)^2 
 + \cdots 
 \,\Biggr] .              
\label{eq:dI_dV}
\end{align}
The result shows that the nonlinear $(eV)^2$ term is also scaled 
by the resonance width $\widetilde{\Delta}$, 
and the coefficient generally depends on 
the parameter $(R-1)^2$, or $\widetilde{U}^2/(\pi \widetilde{\Delta})^2$.
As mentioned, in the strong-coupling limit $u \to \infty$, 
the two characteristic parameters 
become $\widetilde{\Delta} \to (4/\pi)\, T_K$ and    
$R \to 2$.

\subsection{Comparison with other approaches } 

To our knowledge, similar attempts to calculate 
the coefficient $c_V$ of the $(eV)^2$ term of $dI/dV$ have been made 
by two groups\cite{KNG,KSL} in the strong-coupling limit $u \to \infty$:  
\begin{align}
\frac{dI}{dV} \,= \, 
\frac{2e^2}{h} \Biggl[\, 1 - c_T \left(\frac{\pi T}{T_K}\right)^2 
-  c_V \left(\frac{eV}{T_K}\right)^2 + \cdots \Biggr] .
\end{align}
To compare the results, the difference in a numerical factor 
of order $1$ in the definition $T_K$ must be 
taken into account. To avoid this uncertainty, 
we use $T_K$ defined in eq.\ (\ref{eq:T_K}), and rescale 
the results presented in refs.\ \citen{KNG} and \citen{KSL} 
such that the coefficient for the linear-response $T^2$ term 
agrees with the result of Yamada and Yosida $c_T = (\pi/4)^2$. 
Kaminski, Nazarov, and Glazman\cite{KNG} have carried out 
a perturbation expansion 
around the strong-coupling fixed point to 
obtain $c_V^{\rm KNG} = (3/8)\,c_T$.
 Konik, Saleur, and Ludwig\cite{KSL} have used 
 the equilibrium Bethe ansatz solution, 
and then made some extra assumptions for calculating 
the nonlinear coefficient to obtain $c_V^{\rm KSL} = 4\, c_T$ 
(the parameters corresponding to $\alpha_L$ and  $\alpha_R$ 
used by KSL seem to be different from ours).
Our result eq.\ (\ref{eq:dI_dV}), 
which is obtained using the Ward identities,  
shows $c_V^{\rm Ward} = (3/2) \,c_T$ in the strong-coupling limit.

Although the Hamiltonian is somewhat different, 
we also note for comparison that 
Schiller and Hershfield\cite{Schiller} obtained the result 
corresponding to $c_V^{\rm SH} = 3\,c_T$ for a special 
parameter set which can be related to the Emery-Kivelson solution of 
the two-channel Kondo model.

\section{Renormalized Perturbation Theory at Finite Bias Voltages}
\label{sec:RPT}

Although the description of the low-energy properties 
discussed in \S \ref{sec:low-voltage} is exact, 
the underlying physics of the quasiparticles 
might not be seen directly in the microscopic derivation.
In the case of the three-dimensional Fermi liquid,  
the vertex function played a central role to clarify 
a link between the intuitive picture of the quasiparticles 
and Green's functions.
Specifically, 
the residual interaction between two quasiparticles,
which had been introduced phenomenologically, 
was shown to be connected to the forward scattering amplitude.\cite{AGD}

For the Anderson impurity the vertex function at Fermi energy, 
 $z^2 \Gamma_{\uparrow\downarrow;\downarrow\uparrow}(0,0;0,0)$,
corresponds to the scatter amplitude,
and it is equal to $\widetilde{U}$ by the 
 definition in eq.\ (\ref{eq:renorm_parm}). 
The perturbation expansion in $\widetilde{U}$, 
which has been formulated precisely 
in the equilibrium case by Hewson,\cite{HewsonRPT} 
provides the link between the quasiparticles 
and microscopic theory of the local Fermi liquid.
All the basic Fermi-liquid behavior have been shown to be   
reproduced in the expansion up to terms of 
order $\widetilde{U}^2$. Furthermore, 
the approach is not limited to low energies. 
To carry out the expansion systematically, however, 
one has to take account of the renormalization conditions 
that are necessary to avoid overcounting of the many-body effects, 
because in the renormalized perturbation theory (RPT) 
the expansion parameter already contains some 
contributions of the Coulomb interaction. 

In this section we apply  the RPT to the nonequilibrium steady state.
It reproduces the result of $dI/dV$ in the Fermi-liquid regime,
and gives us one possible way to calculate the corrections needed 
at high voltages.
For simplicity, we concentrate on the electron-hole symmetric case; 
$\epsilon_d = -U/2$, $\Gamma_L = \Gamma_R$, and $\alpha_L=\alpha_R=1/2$. 
The unperturbed Green's function is defined 
such that it describes the Kondo resonance with 
the renormalized level width   
$ \widetilde{G}_0^{r}(\omega) = \bigl[\,\omega  
  + \text{i} \widetilde{\Delta}\,\bigr]^{-1}$,
  as that in the equilibrium case.\cite{HewsonRPT} 
%%
%% $\widetilde{G}_0^{a}(\omega) 
%% = \{ \widetilde{G}_0^{r}(\omega) \}^*$. 
%%
However,
in the nonequilibrium case, 
it is not obvious how the distribution function 
for the free quasiparticles should be given by.
We simply assume here that it is given by 
the noninteracting one, which in the electron-hole symmetric case 
takes the form 
$f_{\rm eff}(\omega)=\left[\, f(\omega-eV/2) 
+ f(\omega+eV/2) \,\right]/2$.  
Hence, the four elements of the unperturbed Green's functions, 
$\widetilde{\mbox{\boldmath $G$}}_{0}$, 
take the forms 
\begin{align}
 &\!
 \widetilde{G}_0^{--}(\omega) =  
 \bigl[ 1-f_{\text{eff}}(\omega) \bigr]\,
     \widetilde{G}_0^r(\omega)
      +   f_{\text{eff}}(\omega)\,
      \widetilde{G}_0^a(\omega) ,
\label{eq:G_^--} 
\\
&\!
\widetilde{G}_0^{-+}(\omega) =  
-\,f_{\text{eff}}(\omega)
        \left[\,\widetilde{G}_0^r(\omega)- 
        \widetilde{G}_0^a(\omega)\,\right] , 
\label{eq:G_^-+}
\\
&\!
\widetilde{G}_0^{+-}(\omega) =  
\bigl[ 1-f_{\text{eff}}(\omega) \bigr]
        \left[\,\widetilde{G}_0^r(\omega) 
        - \widetilde{G}_0^a(\omega)\,\right] , 
\label{eq:G_^+-}
\\
 & \!
 \widetilde{G}_0^{++}(\omega) =  
 -\bigl[ 1-f_{\text{eff}}(\omega) \bigr]\,
     \widetilde{G}_0^a(\omega)
      -   f_{\text{eff}}(\omega) \,\widetilde{G}_0^r(\omega) \;.
\label{eq:G_^++}
\end{align}
Correspondingly, 
the full propagator of the quasiparticles,
which includes all effects of $\widetilde{U}$, 
is defined by $\widetilde{\mbox{\boldmath $G$}}(\omega) \equiv z^{-1} 
\mbox{\boldmath $G$}(\omega)$. 
Therefore, in terms of the renormalized quantities, 
the nonequilibrium current, eq.\ (\ref{eq:caroli}), 
is written as  
\begin{align}
 I\,  &  =\,  {2 {\sl e} \over h}  \int_{-\infty}^{\infty} 
 \! \text{d}\omega 
    \left[\, f_L - f_R \, \right] 
 \left[\, -\,\widetilde{\Delta}\, 
 {\rm Im}\, \widetilde{G}^r(\omega) \,\right] \;.
\label{eq:caroli_ren}
\end{align}
The self-energy correction due 
to $\widetilde{U}$ satisfies the Dyson equation of the form   
$ \widetilde{\mbox{\boldmath $\Sigma$}}(\omega)\,  
    \equiv   
 \{\widetilde{\mbox{\boldmath $G$}}_{0}(\omega)\}^{-1} 
 - 
\{\widetilde{\mbox{\boldmath $G$}}(\omega)\}^{-1}$. 
Using eq.\ (\ref{eq:Dyson}), we have 
\begin{align}
\widetilde{\mbox{\boldmath $\Sigma$}}(\omega)\,  
&\, = 
 \{\widetilde{\mbox{\boldmath $G$}}_{0}(\omega)\}^{-1} 
\, - \, z \left[\,\{\mbox{\boldmath $G$}_{0}(\omega)\}^{-1} 
-\mbox{\boldmath $\Sigma$}(\omega)\,\right]
\nonumber \\
&\, = \, 
   z\,\mbox{\boldmath $\Sigma$}(\omega)
 \, + \, (1 - z )  \mbox{\boldmath $\tau$}_3 \,\omega
\;,
\label{eq:self_ren}
\end{align}
where
$\mbox{\boldmath $\tau$}_i$ for $i=1,2,3\,$ is the Pauli matrix.

\begin{figure}[tb]
\leavevmode 
\begin{center}
\begin{minipage}{0.3\linewidth}
\includegraphics[width=1.0\linewidth, 
clip, trim = 0cm 0cm 0cm 0cm]{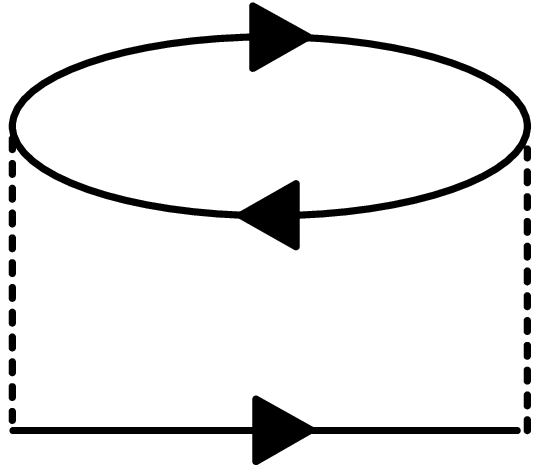}
\end{minipage}
\caption{
The second-order diagram of $\widetilde{\mbox{\boldmath $\Sigma$}}$. 
The dashed represents $\widetilde{U}$, 
and solid lined represents 
the free-quasiparticle propagator 
$\widetilde{\mbox{\boldmath $G$}}_{0}$. 
%% defined by eqs.\ (\ref{eq:G_^--})--(\ref{eq:G_^++}).
} 
\label{fig:2nd_graph}
\end{center}
\end{figure}

\subsection{Low-energy behavior up to terms of order $\omega^2$, $T^2$ 
and $(eV)^2$}

At low-energies,  the $\omega$-linear contributions 
in the right-hand side of eq.\ (\ref{eq:self_ren})  
cancel out owing to eq.\ (\ref{eq:re_S}).
The contributions of order $\omega^2$, $T^2$ and $(eV)^2$    
arise from the second-order diagram 
for $\widetilde{\mbox{\boldmath $\Sigma$}}(\omega)$
shown in Fig.\ \ref{fig:2nd_graph},
where the solid and dotted lines represent  
 $\widetilde{\mbox{\boldmath $G$}}_{0}(\omega)$ and $\widetilde{U}$, 
 respectively. Calculating the contributions from the diagram, 
 and then taking the cancellation of the $\omega$-linear 
 term into account, 
 we have 
\begin{align}
\widetilde{\Sigma}^r(\omega) 
\,= \, -\, \text{i}\,
  {\displaystyle \mathstrut \widetilde{U}^2 
 \over \displaystyle 
 2 \widetilde{\Delta} 
 \rule{0cm}{0.4cm} \mathstrut 
       (\pi \widetilde{\Delta})^2}
\left[ \, \omega^2 + 
{\displaystyle \mathstrut 3  \over \displaystyle \mathstrut 4}
\,({\sl e}V)^2 +(\pi T)^2
       \right] \,+\, \cdots .
\label{eq:self_U2ren}
\end{align}
It simply reproduces the renormalized Green's function 
 $\widetilde {G}^r(\omega)$ corresponding 
 to eq.\ (\ref{eq:Gr_symmetric_case}). 
Furthermore, using eq.\ (\ref{eq:caroli_ren}), 
the $(eV)^2$ term of $dI/dV$ in eq.\ (\ref{eq:dI_dV}) 
is also reproduced in the expansion up to terms 
of order $\widetilde{U}^2$.
Since eqs.\ (\ref{eq:Gr_symmetric_case}) is asymptotically exact, 
the higher-order terms in $\widetilde{U}$ do not 
change the low-energy behavior in eq.\ (\ref{eq:self_U2ren}).

Note that eq.\ (\ref{eq:self_U2ren}) follows from 
the fact that we have used $f_{\text{eff}}(\omega)$ for 
the distribution function of the free quasiparticles.
This assumption seems to be justified also from 
the fact that the many-body effects on 
 the correlated distribution $f_{\text{eff}}^U(\omega)$ 
defined in eq.\ (\ref{eq:f_int}) arise first in 
the order $U^2$ contribution.

\subsection{Beyond the Fermi-liquid regime}

To study the higher-energy behavior 
at large $\omega$, $T$, and $eV$ with the RPT, one needs to calculate 
the higher order terms in $\widetilde{U}$.   
In the following, we describe the outline of the procedure of the expansion. 

At high energies the renormalization factor $z$ cannot be defined 
 with respect to $T=0$ and $eV=0$ no longer.
 This is because the coefficient of the $\omega$-linear term  
of the self-energy  depends on $T$ and $eV$.
 For instance, in the next order, the terms of 
the form $T^2\omega$ and $(eV)^2\omega$ exist. 
Therefore,  $z$ is redefined such that 
the $\omega$-linear contributions in eq.\ (\ref{eq:self_ren})
cancel out
\begin{align}
\biggl.
{
\partial 
\widetilde{\Sigma}^r(\omega) \over 
\partial 
\omega }
\biggr|_{\omega=0} \,=\,0 \;.
\label{eq:norm_condition1}
\end{align}
Hence $z$ generally depends on  $T$ and $eV$.
The perturbation expansion in $\widetilde{U}$ can be carried out 
following that in the equilibrium case.\cite{HewsonRPT}
We first of all rewrite the action $S$  in the form
\begin{align} 
S\, &= \, z^{-1} \widetilde{S}_0 
\,+ \, z^{-2} S_U  \widetilde{U}/\,U
\,-\, S_{\rm cou}
\label{eq:action_rewritten}
\\
S_{\rm cou} &= \, 
\lambda\, z^{-2} S_U \widetilde{U} /\,U 
\, + \,  
 z^{-1} \widetilde{S}_0 - S_0 
\;,
\label{eq:action_counter}
\end{align} 
where $\widetilde{S}_0$ is the action for 
the free quasiparticle  
corresponding to 
the propagator $\widetilde{\mbox{\boldmath $G$}}_{0}(\omega)$, and   
$\lambda \equiv 1 - z^2\,U/\widetilde{U}$. 
In eqs.\ (\ref{eq:action_rewritten}) 
and (\ref{eq:action_counter}) 
the factor $1/U$ is introduced just to cancel 
the bare Coulomb interaction $U$ included in $S_U$ by 
the definition in eq.\ (\ref{eq:action_U}). 
The perturbation series in $\widetilde{U}$ is generated
by taking  $z^{-1}\widetilde{S}_0$ in eq.\ (\ref{eq:action_rewritten})
to be the unperturbed part 
and taking the remaining terms 
$ z^{-2} S_U  \widetilde{U}/U - S_{\rm cou}$ to be the perturbed part.
Here, $S_{\rm cou}$ is 
the counter-term which avoids overcounting of the many-body effects. 
Specifically, the last two terms in the right-hand side of 
eq.\ (\ref{eq:action_counter}), which can be rewritten in the form
 \begin{align}
 z^{-1}\widetilde{S}_0  - S_0 
 &=   \left(z^{-1} - 1\right)
    \sum_{\sigma} \int_{-\infty}^{\infty} \!\! \text{d}\omega\, 
 \mbox{\boldmath $\eta$}_{\sigma}^{\dagger}(\omega) 
 \mbox{\boldmath $\tau$}_3 \,\omega\,
 \mbox{\boldmath $\eta$}_{\sigma}(\omega) \,. 
 \end{align}
It corresponds to the counter-term for the renormalization factor $z$.
In the RPT, the two parameters $z$ and $\lambda$ are regarded as 
functions of the renormalized parameters 
 $\widetilde{\Delta}$ and $\widetilde{U}$, 
and are expanded as series in the powers of $\widetilde{U}$. 
Then the expansion coefficients for $z$ and $\lambda$ 
can be determined such that the two renormalization conditions,
eqs.\ (\ref{eq:norm_condition1}) and (\ref{eq:norm_condition2}), 
are satisfied by each order in $\widetilde{U}$; 
\begin{align}
%% \left.
%% {
%% \partial 
%% \widetilde{\Sigma}^r(\omega) /
%% \partial 
%% \omega }
%% \right|_{\omega=0} \,=\,0 \;, \qquad
\widetilde{\Gamma}_{\uparrow\downarrow;\downarrow\uparrow}(0,0;0,0) 
\,= \,\widetilde{U} \;.
\label{eq:norm_condition2}
\end{align}
Here  $\widetilde{\Gamma}_{\uparrow\downarrow;\downarrow\uparrow}
\equiv z^2 \,\Gamma_{\uparrow\downarrow;\downarrow\uparrow}
$ is the vertex part for the four external causal Green's 
functions $\widetilde{G}^{--}$.
In the RPT, 
$\widetilde{\Gamma}_{\uparrow\downarrow;\downarrow\uparrow}(0,0;0,0)$
is calculated in the power series in $\widetilde{U}$, 
and at high-energies it generally depends on $T$ and $eV$.
Note that the contribution of the parameter $\lambda$ first arises 
in the order $\widetilde{U}^3$ terms.\cite{HewsonRPT} 
For this reason, the condition 
of $\lambda$ is not necessary to be taken into account in the expansion 
up to order $\widetilde{U}^2$.

As already mentioned, higher-order terms in $\widetilde{U}$ 
are needed to study the high-energy behavior of 
 $\widetilde{\Sigma}^r(\omega)$ and $dI/dV$ 
beyond the $\omega^2$- and $(eV)^2$-terms. 
One possibility is to include the contributions 
up to terms of order $\widetilde{U}^4$. 
The corresponding calculations in the bare-$U$ expansion 
have been carried out by Fujii and Ueda.\cite{FujiiUeda} 
Alternatively, in the equilibrium case at $eV=0$,
a combination of the RPT and NRG has been examined recently,
and the results reproduce the $T$-dependence of 
the spin susceptibility accurately 
in a wide temperature range.\cite{HewsonNRG}
Such a combination would be another possibility to go beyond 
the Fermi-liquid regime at large bias voltages. 
%%
%%  

%% \medskip

\section{Summary}
\label{sec:SUMMARY}

 We have studied the low-energy properties of 
 the Anderson model under a finite bias voltage $V$ 
 using the properties of 
 the Keldysh Green's function at the impurity site
  $\mbox{\boldmath $G$}(\omega)$  
 as a functional of the nonequilibrium distribution 
 function $f_{\text{eff}}(\omega)$.
 Through the distribution function $f_{\text{eff}}(\omega)$,
 the $T$- and $eV$-dependence of $\mbox{\boldmath $G$}(\omega)$ arise. 
  The Ward identities for the derivative of the self-energy
  with respect to $eV$ follow from these properties 
  that can be summarized in the form of eq.\ (\ref{eq:functional_f1}).
 Using the Ward identities, 
 the differential conductance $dI/dV$  has been 
 determined up to terms of order $(eV)^2$ in eq.\ (\ref{eq:dI_dV}) 
 in the electron-hole symmetric case. 
 The coefficients are determined by 
 two characteristic parameters $\widetilde{\Delta}$ and $R$.

 We have also described the low-energy properties using 
 the renormalized perturbation theory in the Keldysh formalism.
  To second order in $\widetilde{U}$,   
 it reproduces the exact $(eV)^2$ coefficients for $dI/dV$. 
 The Fermi-liquid behavior 
 of $dI/dV$ follows from the assumption that the distribution function for 
 the free quasiparticles are the same as that of the noninteracting 
 electrons $f_{\text{eff}}(\omega)$.  
 In order to study the corrections to the Fermi liquid theory 
 at large bias voltages with the RPT, 
 one needs to calculate the higher order terms in $\widetilde{U}$.

%% \bigskip

\section*{Acknowledgements}

I wish to thank A. C. Hewson for helpful discussions on 
the renormalized perturbation theory.  
This work was supported 
by the Grant-in-Aid for Scientific Research from JSPS.

%% \bigskip

\appendix
 
\section{Derivation of 
eqs.\ (\ref{eq:derivative_self_1})--(\ref{eq:derivative_self_2_with_CII})}

The first derivative of the self-energy with respect to $eV$ can be 
written, using eq.\ (\ref{eq:functional_f1}), as
%% \begin{full}
\begin{align}
& \biggl.
{
\partial \mbox{\boldmath $\Sigma$}_{\sigma}(\omega)
\over
\partial (eV)}
\biggr|_{eV=0}
\ = 
 \nonumber \\
&   
\qquad  \sum_{\nu\nu'\sigma'} 
\int_{-\infty}^{\infty} \! {\text{d}\omega'}   
{
\delta \mbox{\boldmath $\Sigma$}_{{\rm eq},\sigma}(\omega)
\over
\delta G_{0:{\rm eq},\sigma'}^{\nu\nu'}(\omega')
}\,
\biggl.
  {\partial  
 G_{0\sigma'}^{\nu\nu'}(\omega') 
 \over
 \partial (eV)} 
 \biggr|_{eV=0}
 .
\label{eq:functional_f1v}
\end{align}
%% \end{full}
%%
Similarly, the derivative 
of $\mbox{\boldmath $\Sigma$}_{\sigma}(\omega)$ with 
respect to $\omega$ at $eV=0$ is written in the form, 
\begin{align}
& \!\!\!\!\!\!\!\!\!\!\!\!\!
\left(
\frac{\partial}{\partial \omega}
+
\frac{\partial}{\partial E_d}
\right)
\mbox{\boldmath $\Sigma$}_{{\rm eq},\sigma}(\omega)
 \nonumber \\
& = \   
 \sum_{\nu\nu'\sigma'} 
\int_{-\infty}^{\infty} \! {\text{d}\omega'}   
{
\delta \mbox{\boldmath $\Sigma$}_{{\rm eq},\sigma}(\omega)
\over
\delta G_{0:{\rm eq},\sigma'}^{\nu\nu'}(\omega'+\omega)
}
 \nonumber \\
 & \qquad \qquad \qquad \times 
\left(
\frac{\partial}{\partial \omega}
+
\frac{\partial}{\partial E_d}
\right)
 G_{0:{\rm eq},\sigma'}^{\nu\nu'}(\omega'+\omega) 
\nonumber \\
& =  \  
 \sum_{\nu\nu'\sigma'} 
\int_{-\infty}^{\infty} \! {\text{d}\omega'}   
{
\delta \mbox{\boldmath $\Sigma$}_{{\rm eq},\sigma}(\omega)
\over
\delta G_{0:{\rm eq},\sigma'}^{\nu\nu'}(\omega')
}
 \nonumber \\
 & \qquad \qquad \qquad   \times 
\left(
\frac{\partial}{\partial \omega'}
+
\frac{\partial}{\partial E_d}
\right)
 G_{0:{\rm eq},\sigma'}^{\nu\nu'}(\omega') 
 .
\label{eq:functional_f1w}
\end{align}
Here we have used the property  
that the frequency $\omega'$ can be shifted to $\omega'+\omega$ 
without changing the result. 
This is because the value of the self-energy does not change 
if all the frequencies which are assigned to the Green's functions 
in a closed-loop diagram are shifted by the same amount.
Therefore,  substituting eq.\ (\ref{eq:derivative_1b}) 
into eq.\ (\ref{eq:functional_f1v}), 
we obtain eq.\ (\ref{eq:derivative_self_1}).

To calculate the second derivative,
the variation of the self-energy 
$\delta \mbox{\boldmath $\Sigma$}_{\sigma}$ 
must be calculated up to terms 
of order $(\delta \! f_{\text{eff}})^2$, and 
we find
\begin{align}
& \biggl.
{
\partial^2 \mbox{\boldmath $\Sigma$}_{\sigma}(\omega)
\over
\partial (eV)^2}
\biggr|_{eV=0}
 \nonumber \\
& = \  \sum_{\nu\nu'\sigma'} 
\int_{-\infty}^{\infty} \! {\text{d}\omega'}   
{
\delta \mbox{\boldmath $\Sigma$}_{{\rm eq},\sigma}(\omega)
\over
\delta G_{0:{\rm eq},\sigma'}^{\nu\nu'}(\omega')
}\,
\biggl.
  {\partial^2  
 G_{0\sigma'}^{\nu\nu'}(\omega') 
 \over
 \partial (eV)^2} 
 \biggr|_{eV=0}
\nonumber\\
& \ \ \  +
 \sum_{\scriptstyle \nu_1 \nu_2, \sigma' 
 \atop \scriptstyle \nu_3 \nu_4, \sigma''} 
\!\int_{-\infty}^{\infty} \! 
{\text{d}\omega'}   {\text{d}\omega''}   
{
\delta^2 \mbox{\boldmath $\Sigma$}_{{\rm eq},\sigma}(\omega)
\over
\delta G_{0:{\rm eq},\sigma'}^{\nu_1 \nu_2}(\omega')
\delta G_{0:{\rm eq},\sigma''}^{\nu_3 \nu_4}(\omega'')
}\,
\nonumber \\
& \qquad \qquad \qquad \qquad \times  {\partial  
 G_{0\sigma'}^{\nu_1\nu_2}(\omega') 
 \over
 \partial (eV)} \,
\biggl.
  {\partial  
 G_{0\sigma''}^{\nu_3 \nu_4}(\omega'') 
 \over
 \partial (eV)} 
 \biggr|_{eV=0}
 .
  \end{align}
Then flowing along the similar line, 
we obtain eq.\ (\ref{eq:derivative_self_2_with_CII}) 
using eq.\ (\ref{eq:derivative_1b}) and    
the corresponding relation for the second derivative 
\begin{align}
\left.
{\partial^2  G_0^{\nu\nu'}(\omega)
\over
\partial ({\sl e}V)^{2}} \right|_{eV=0}
&= \,
\kappa 
\left(
{\partial \over \partial \omega}
+ {\partial \over \partial E_d}
\right)^{2}
G_{0:\text{eq}}^{\nu\nu'}(\omega)\;,
\end{align}
where
$
\kappa \equiv 
     {
  ( \alpha_L^2 \Gamma_L + \alpha_R^2 \Gamma_R ) / 
  ( \Gamma_L+ \Gamma_R )} 
$.

\end{document}